\documentclass[conference]{IEEEtran}
\usepackage{listings}
\usepackage{tikz}

\usepackage{epsfig,endnotes}
\usepackage{multirow}

\usepackage{url}

\usepackage{color}
\usepackage{url}

\usepackage{hyperref}
\usepackage{breakurl}

\usepackage{amsmath,amssymb,amsfonts}
\usepackage{algorithmic}
\usepackage{graphicx}
\usepackage{textcomp}
\usepackage{xcolor}
\usepackage{pifont}
\usepackage{ulem}
\usepackage{subfigure}

\definecolor{codeauburn}{rgb}{0.43, 0.21, 0.1}
\definecolor{codegreen}{rgb}{0,0.6,0}
\definecolor{codegray}{rgb}{0.5,0.5,0.5}
\definecolor{codepurple}{rgb}{0.58,0,0.82}
 
\lstdefinelanguage
   [x64]{Assembler}     
   [x86masm]{Assembler} 
   {morekeywords={CDQE,CQO,CMPQ,CMPXCHG16B,JRCXZ,LODSQ,MOVSXD,movl,addl,subl, %
                  POPQ,PUSHQ,SCASQ,STOSQ,IRETQ,RDTSCP,SWAPGS,MOVABSQ,LEAQ, %
                  rax,rdx,rcx,rbx,rsi,rdi,rsp,rbp,retq,callq,movq, %
                  r8,r8d,r8w,r8b,r9,r9d,r9w,r9b,r10,r11,r12,r15,enclu,rdgsbase,wrgsbase,ud2,mfence,lfence,cmova,cmovg,cmovl}} 
                  
\lstdefinestyle{mystylenobox}{
    backgroundcolor=\color{white},   
    commentstyle=\color{blue},
    keywordstyle=\color{codeauburn},
    numberstyle=\tiny\color{codegray},
    stringstyle=\color{codepurple},
    basicstyle=\footnotesize,
    breakatwhitespace=false,         
    breaklines=true,                 
    captionpos=b,                    
    keepspaces=true,                 
    numbers=left,                    
    numbersep=-1pt,
    xleftmargin=.25in,
    showspaces=false,                
    showstringspaces=false,
    showtabs=false, 
    tabsize=2
}

\makeatletter
\let\@ORGmakecaption\@makecaption
\long\def\@makecaption#1#2{\@ORGmakecaption{#1}{#2}\vskip\belowcaptionskip\relax}
\makeatother

\newcounter{packednmbr}
\newenvironment{packedenumerate}{
\begin{list}{\thepackednmbr.}{\usecounter{packednmbr}
\setlength{\itemsep}{0pt}
\addtolength{\labelwidth}{3pt}
\setlength{\leftmargin}{19pt}
\setlength{\listparindent}{0.8\parindent}
\setlength{\parsep}{1pt}
\setlength{\topsep}{1pt}}}{\end{list}}

\usepackage{amsthm}

\usepackage{listingsutf8}
\definecolor{mulberry}{rgb}{0.772,0.29,0.549}

\newcommand{\ignore}[1]{}
\newcommand\weijie[1]{\textcolor{brown}{\{\textbf{weijie:} {\em#1}\}}}
\newcommand\hongbo[1]{\textcolor{teal}{\{\textbf{hongbo:} {\em#1}\}}}

\IEEEoverridecommandlockouts

\date{}

\begin{document}

\date{}

\title{Understanding TEE Containers, Easy to Use? Hard to Trust}

\author{\IEEEauthorblockN{Weijie Liu\IEEEauthorrefmark{1}\textsuperscript{\textsection},
Hongbo Chen\IEEEauthorrefmark{1}\textsuperscript{\textsection},
XiaoFeng Wang\IEEEauthorrefmark{1},
Zhi Li\IEEEauthorrefmark{2},
Danfeng Zhang\IEEEauthorrefmark{3},
Wenhao Wang\IEEEauthorrefmark{5},
Haixu Tang\IEEEauthorrefmark{1}
}

\IEEEauthorblockA{\IEEEauthorrefmark{1}Indiana University Bloomington
\{weijliu, hc50, xw7, hatang\}@iu.edu}
\IEEEauthorblockA{\IEEEauthorrefmark{2}School of Cyber Science and Engineering,\\
Huazhong University of Science and Technology, China
\{lizhi16\}@hust.edu.cn}
\IEEEauthorblockA{\IEEEauthorrefmark{3}Pennsylvania State University
\{dbz5017\}@psu.edu}
\IEEEauthorblockA{\IEEEauthorrefmark{5}Institute of Information Engineering, CAS
\{wangwenhao\}@iie.ac.cn}

}

\maketitle

\begingroup\renewcommand\thefootnote{\textsection}
\footnotetext{Both the first two authors contributed equally to this research.}
\endgroup

\thispagestyle{empty}

\begin{abstract}

As an emerging technique for confidential computing, trusted execution environment (TEE) receives a lot of attention. 
To better develop, deploy, and run secure applications on a TEE platform such as Intel's SGX, both academic and industrial teams have devoted much effort to developing reliable and convenient TEE containers. In this paper, we studied the isolation strategies of 15 existing TEE containers to protect secure applications from potentially malicious operating systems (OS) or untrusted applications, using a semi-automatic approach combining a feedback-guided analyzer with manual code review. Our analysis reveals the isolation protection each of these TEE containers enforces, and their security weaknesses\ignore{ in their design and implementations}. We observe that none of the existing TEE containers can fulfill the goal they set, due to various pitfalls in their design and implementation. We report the lessons learnt from our study for guiding the development of more secure containers, and further discuss the trend of TEE container designs.  We also release our analyzer that helps evaluate the container middleware both from the enclave and from the kernel.   \looseness=-1



\end{abstract}

\section{Introduction}

In response to the growing demands for scalable data protection in computing, the recent decade has witnessed the rapid advance in \textit{trusted execution environment} (TEE) technologies, with the focus on CPU-based isolated execution.  Prominent examples include Intel's Software Guard eXtensions (SGX), AMD's Secure Encrypted Virtualization (SEV) and ARM's TrustZone. Some of these technologies, particularly SGX and TrustZone, have already been widely deployed, protecting real-world computing tasks ranging from password protection~\cite{krawiecka2018safekeeper}, SSL~\cite{han2017sgx} to various cloud services~\cite{azure_sql_doc}. Continued adoption of these technologies, however, can be impeded by their limited supports for unmodified programs.  Particularly, a user of SGX is expected to incorporate its SDK into her original program to execute it inside an SGX enclave, a process that can entail a lot of effort. To a lesser extent, code to be run inside SEV and TrustZone may also require modification to work under the VM and OS supported by the hardware.  Addressing this usability challenge are a set of container-style TEE middleware, such as Graphene-SGX~\cite{tsai2017graphene}, SCONE~\cite{arnautov2016scone}, etc., which enable either direct running of unmodified binary code inside a TEE or automated transformation of source code before loading it into a TEE executable. In our research we call such middleware \textit{TEE container} or simply \textit{Tcon}.  A Tcon is meant to facilitate use of TEE but it also increases the complexity of the TEE software stack. Less known is whether they will undermine the protection promised by TEEs, a problem that has never been systematically studied before.


\vspace{3pt}\noindent\textbf{Understanding Tcons}. Given the importance of Tcons to TEE-based \textit{confidential computing}, understanding their security properties is of critical importance. To this end, we conducted the \textit{first systematic study} on these Tcons.  Our study focuses on the Tcons for SGX, as they are the mainstay of today's TEE middleware.
In our research, we first survey 15 popular Tcons (Occlum~\cite{shen2020occlum} and its artifact evaluation version are considered as two Tcons), using the information from public sources (research papers, developer guides, source code, etc.), and then propose a taxonomy based upon a set of security properties expected from those containers. These properties include threat models (particularly whether the application \textit{inside} an enclave can be trusted), size of TCBs, the isolation techniques to protect interfaces with the untrusted OS (API, interruption surfaces in particular) and to separate untrusted in-enclave code from sensitive data, side-channel protection, remote attestation support and secure storage mechanism.   

Our survey study on the most popular Tcons, has brought to light some security properties they claim to have and their underlying techniques. However, still less clear is whether these properties are indeed achieved by these techniques and 
whether protection is in place for other properties that have not been publicly stated.  For example, the documentations of most Tcons fail to mention their protection on the APIs, including parameters delivered to the OS (which could lead to information leaks) and values returned from the OS (which could cause security risks such as an Iago attack~\cite{checkoway2013iago}). Another issue is the gap between design and implementation. For example, a quick scan on the source code of Deflection reveals that the covert protection claimed in its paper~\cite{liu2021practical} does not seem to be implemented. Most concerning here is whether these containers can still ensure isolated execution, which is at the center of TEE's security guarantee. The isolation here includes the protection of Tcon-OS interfaces (Ecall, Ocall, exceptions), in-enclave separation (e.g., use of software-based fault isolation) and side-channel control.  Understanding the effectiveness of such protection requires an in-depth analysis on individual Tcons, which so far has not been done.

\vspace{3pt}\noindent\textbf{Analyzing Tcon isolation}. To demystify isolated execution in popular Tcons, we designed a methodology for systematic evaluation of all the interfaces that need to be protected.  Our methodology relies on an automated analyzer, called \textit{TEE Container Fuzzer} or \textit{TECUZZER}, to check most attack surfaces, including Ocall, exception, in-enclave separation and common side channels, and manual code inspection to analyze the Ecall surfaces (which involves understanding of code semantics, such as the presence and proper implementation of remote attestation).  Of particular interest here is the design and implementation of TECUZZER, which is meant to bridge the semantic gap between a program inside a Tcon and the OS outside, so as to evaluate the protection offered by the middleware in-between. 
For example, a Tcon's control on the system call (Ocall) surface can only be analyzed by issuing calls from its inside and then observing the call requests outside, in the OS. For this purpose,  we designed a unique \textit{2-piece} fuzzer, with one component running inside a Tcon within an enclave and the other component operating inside the OS kernel. These two components work together to evaluate the protection implemented by the Tcon using the test case going through the middleware.


More specifically, TECUZZER is designed to fuzz the syscall (Ocall) interface to understand the target Tcon's control on call parameters and return values, to evaluate the in-Tcon protection (for the container not trusting the application it runs) using a set of isolation rules proposed in prior studies, and test the Tcon's side-channel surfaces with proof-of-concept attack instances. Particularly, in order to evaluate a Tcon's handling of syscall parameters and return values, we designed TECUZZER to not only generate random values but also automatically produce those semantically correct ones, which more likely will penetrate the container, revealing potentially dangerous  interactions between application inside and the untrustred OS outside. Our use of Rust enables convenient and effective inspection on the parameter and value types both inside a Tcon and the kernel. The design of TECUZZER also manages the stateful syscalls and correlated parameters and returns.  \looseness=-1




\vspace{3pt}\noindent\textbf{Findings and takeaways}. Running TECUZZER on existing Tcons, we gained new knowledge on how well these Tcons enforce isolated execution as they promise. 
More specifically, our analysis reveals the set of important syscalls being regulated by each Tcon, which have never been made public. Some of these regulated syscalls are handled at the container layer, thereby avoiding exposure of the applications they host to the untrusted OS. For the syscalls that need to be processed by the OS, Graphene-SGX, Occlum and SGX-LKL perform sanitization on the call parameters and return values. However, such protection turns out to be inadequate, as it still leaves the door open to both Iago attack and covet channel leaks when the application inside the containers are untrusted (e.g., a 3MBps channel can be established using the \texttt{nanosleep} syscall and a kernel hook across Deflection).  Also discovered in our study is the weak in-Tcon isolation through software-based fault isolation (SFI): particularly,  containers with the SFI protection (Chancel and Deflection) fail to mediate direct jumps; as a result, all guards put in place can be circumvented. Moreover, although all Tcons claim side-channel attacks are outside their threat models, our analysis shows that Graphene-SGX actually implements some protection against interrupt-based attacks~\cite{he2018sgxlinger}. Finally, by inspecting the Ecall interfaces of existing Tcons, we found that their Ecall interfaces are not well guarded, and hence, they might import untrusted data and code. Further, some of the containers do not implement a complete attestation primitive or fail to implement it at all: particularly, some of the Tcons do not provision secret for encrypting their file systems, so the sensitive data stored there are exposed.  \looseness=-1

Our research shows that Tcon developers do not provide full information about what security properties their containers offer, and where they fall short. As a consequence, the Tcons may not be properly used to build secure services. To confirm our hypothesis, we inspected the use cases of Tcons as published at prominent security venues. Indeed we found that a proposed Intrusion Detection System~\cite{kuvaiskii2018snort} inside Graphene-SGX ignores the fact that the timing provided by the Tcon is from the untrusted OS and can therefore be misled; also an SGX-enforcing gateway~\cite{schwarz2020seng} implemented in Graphene-SGX can also be undermined by the untrusted information from the untrusted OS: e.g., the formats of traffic records read into the container are not verified and DNS records are acquired from untrusted sources. Most importantly, our research shows that Tcon developers should tighten up the control on the interfaces critical for isolated execution, and also fully explain to their users the limitations of their protection, so additional defense can be built into the application layer.  The detailed findings of our study are made available online~\cite{code_release}.


\vspace{3pt}\noindent\textbf{Contributions}.
The paper's contributions are outlined below:

\vspace{2pt}\noindent$\bullet$\textit{ Survey and taxonomy}. We report the first survey study on TEE containers and propose a taxonomy for categorizing these systems and understanding their security protection. 

\vspace{2pt}\noindent$\bullet$\textit{ New Tcon analysis technique}. We built a new Tcon analyzer that utilizes both an in-container component and an out-container kernel module to jointly evaluate the security of Tcon middleware. The development of the analyzer addressed technical and engineering challenges.  It is the first tool with such capabilities that has been made publicly available~\cite{code_release}.

\vspace{2pt}\noindent$\bullet$\textit{  New understanding and suggestions}. Our study has led to new and surprising findings about the isolation protection built into popular TEE containers.  The design pitfalls and implementation lapses identified will help enhance Tcon's security quality, which is critical to the wide adoption of TEE-based confidential computing. 

\section{Survey on Mainstream TEE Containers}
\label{sec:background}


\begin{table}[!htbp]
\belowcaptionskip=-10pt
\caption{Summary of TEE Containers}
\label{tab:middleware-category}
\begin{center}
\begin{tabular}{|c|c|c|c|}
\hline
Tcons & Threat model & Inner isolation       & AMI   \\ \hline
Graphene-SGX   & T            & N/A                      & LibOS                            \\ \hline
SCONE      & T            & N/A & musl-libc                           \\ \hline
Occlum     & T          & N/A                      & LibOS                               \\ \hline
Occlum AE     & U          & SFI                      & LibOS                            \\ \hline
SGX-LKL    & T            & N/A                      & LibOS                             \\ \hline
Chancel    & U            & SFI                        & musl-libc/tlibc                 \\ \hline
Deflection & U            & SFI                        &  musl-libc                      \\ \hline
Ratel      & T            & DBT                      & musl-libc/glibc                    \\ \hline
Ryoan      & U            & SFI (NaCl)                       & eglibc                     \\ \hline

MesaPy      & T            & N/A                      & RPython                 \\ \hline
EGo        & T            & N/A                  & glibc       \\ \hline
GOTEE      & U            & N/A                       & Own runtime                   \\ \hline
AccTEE      & U            & Sandbox                       & Own interpreter              \\ \hline
TWINE      & T            & Sandbox       & WASI            \\ \hline
Enarx      & T            & Sandbox             & WASI        \\ \hline

\end{tabular}

T: code running inside the enclave is trusted.

U: code running inside the enclave is untrusted.
\end{center}
\vspace{-15pt}
\end{table}


A TEE \textit{middleware} supports running of applications inside a TEE. For example, Intel provides an SGX SDK~\cite{sgxsdk} as middleware for developing enclave programs; other prominent SGX SDKs include Google Asylo~\cite{asylo_official},  Microsoft Open Enclave~\cite{openenclave_official}, Teaclave SGX/Trustzone SDK~\cite{wang2019towards, wan2020rustee}, and Edgeless RT~\cite{edgelessrt_repo}. Such SDK-based middleware needs to be integrated into a TEE program and therefore, requires considerable manual effort to transform legacy code to utlize the security features of TEE. To avoid this limitation, \textit{TEE containers} are developed to enable direct running of legacy binary or automated conversion and compilation of legacy source code into a TEE executable. Such a container can be loosely considered to be a special case of cloud-based container virtualization, such as Docker~\cite{docker}, which includes all dependencies (e.g., packages, libraries, and other binaries) for running an application.  With some of these Tcons claiming to support 
AMD SEV (Kata Container~\cite{katacontainer}), most Tcons today are built for Intel SGX, given the popularity of the TEE platform and the challenges in developing its enclave programs. Therefore, we focus in our research on SGX Tcons. 

To analyze these TEE containers, we propose a taxonomy including a set of key security properties expected from a Tcon: 1) their threat models, 2) their supports for isolation between the untrusted OS and the container (through Ecall/Ocall/exception interfaces), 3) their supports for isolation within the container (particularly for those running untrusted code), 4) their mechanism for attestation, 5) protection for storage and 6) their side-channel control. These properties are summarized from Tcon-related publications~\cite{tsai2017graphene, wang2020towards, arnautov2016scone, priebe2019sgx, liu2021practical, goltzsche2019acctee, menetrey2021twine, ghosn2019secured, cui2021dynamic, shen2020occlum, ahmad2021chancel, hunt2018ryoan} and documentations~\cite{graphene_doc, occlum_doc, scone_doc, enarx_doc}.
In the rest of the section, we first present popular Tcons and their backgrounds, and then analyze them using the taxonomy.  \looseness=-1

\subsection{Existing TEE Containers}
\label{subsec:existingtcons}

In this paper, we use TEE containers to refer to TEE middleware that are meant to either directly run unmodified legacy binary, or automatically compile unmodified legacy source code into a TEE executable. Next, we introduce existing Tcons in these two categories. 

\vspace{3pt}\noindent\textbf{Tcons hosting unmodified binary}. Tcons hosting unmodified executables include Graphene-SGX, Occlum, SGX-LKL, and Ratel. \looseness=-1



\vspace{3pt}\noindent$\bullet$\textit{ Graphene-SGX}. Graphene-SGX is an open-source library OS (LibOS)~\cite{porter2011rethinking} to run unmodified Linux binaries inside an SGX enclave~\cite{tsai2017graphene}; it transparently handles all interactions cross the enclave boundary. Specifically, the LibOS offers a limited Ecall interface to launch the application, and serves the system calls made by the shielded application in the enclave or forwards the calls to the untrusted OS. Exceptions and signals are also trapped outside the enclave and forwarded back to the LibOS for secure handling, allowing it to interpose on native syscalls. While Graphene-SGX was originally developed as a research project, it has seen increasing industry adaption and thrives to become a standard Tcon solution in the Intel SGX ecosystem. \looseness=-1

\vspace{3pt}\noindent$\bullet$\textit{ Occlum}.
Occlum is the first memory-safe, multi-processing LibOS for SGX. Since Occlum is written in Rust, it is much less likely to contain low-level, memory-safety bugs and therefore is more trustworthy to host security-critical applications. Occlum supports lightweight multitasking LibOS processes that share the same SGX enclave. Its artifact evaluated (AE) version~\cite{shen2020occlum, occlum_asplos20} implements the Intel MPX-based SFI to prevent memory attacks against untrusted applications. With MPX being depreciated recently~\cite{intelmanual}, the Tcon has moved away from internal isolation, and its current version~\cite{occlum_repo} can no longer host untrusted applications.

\vspace{3pt}\noindent$\bullet$\textit{ SGX-LKL}. SGX-LKL is an open-source research project~\cite{sgx_lkl_repo} that offers a trusted in-enclave LibOS to run unmodified Linux binaries~\cite{priebe2019sgx}. Its LibOS layer is internally based on the Linux Kernel Library (LKL). Note that a variant of SGX-LKL can provide in-enclave isolation~\cite{purdila2010lkl}.

\vspace{3pt}\noindent$\bullet$\textit{ Ratel}.
Ratel~\cite{cui2021dynamic} is a new framework that enables dynamic binary translation on SGX. It offers complete interposition, the ability to interpose on all executed instructions in the enclave and monitor all interactions with the OS, by porting a Dynamic Binary Translation engine (DynamoRIO) into enclave. The design of Ratel chooses completeness in its interposition over performance, whenever conflicts arise.


\vspace{3pt}\noindent\textbf{Tcons recompiling unmodified source code}. Most Tcons compile legacy C and Rust source code into TEE executables, with some systems working on GO and Javascript code or running WASM (WebAssembly) executables.  


\vspace{3pt}\noindent$\bullet$\textit{ SCONE}.
SCONE~\cite{arnautov2016scone} is a Tcon that can compile unmodified source code into an enclave application binary using an SGX-aware musl-libc and/or run unmodified Alpine Linux binary. Unlike some other Tcons, SCONE provides comprehensive encryption protection not only for files, but also for environment variables and input parameters~\cite{scone_arg_doc}. 


\vspace{3pt}\noindent$\bullet$\textit{ Ryoan}.
Ryoan is a distributed sandbox service that leverages SGX to protect sandbox instances from potentially malicious computing platforms~\cite{ryoan16osdi}. The protected instances confine untrusted data-processing modules to prevent leakage of the user’s input data. Ryoan is based on Native Client~\cite{sehr2010adapting,yee2009native}, which uses compiler techniques to confine code. Binaries are checked at the load time to ensure that they are properly restricted. Confined code relies on the sandbox for all interactions with the outside world.

\vspace{3pt}\noindent$\bullet$\textit{ Chancel}.
Chancel\footnote{Chancel's source code is not publicly available. It's still under development and not ready for release.} is a sandbox designed for
multi-client isolation within a single SGX enclave~\cite{ahmad2021chancel}. It allows a program’s threads to access both a per-thread memory region and a shared read-only memory region while servicing requests. Each thread handles requests from a single client at a time and is isolated from other threads, using an SFI scheme.

\vspace{3pt}\noindent$\bullet$\textit{ Deflection}. Deflection~\cite{liu2021practical,deflection_code} provides practical and efficient in-enclave code verification, which allows the user to check some security policies of the code provided by untrusted parties. It is also a full-stack middleware that can host binary code generated from its compiler toolchain.

\vspace{3pt}\noindent$\bullet$\textit{ MesaPy}. MesaPy~\cite{wang2020towards, mesapy_repo} aims to support Python code in SGX enclave. It ports PyPy~\cite{pypy_official} into SGX for executing Python with several popular libraries, and formal verification has been performed against its C code to check memory safety problems. MesaPy eliminates potential I/O operations while running and preloading all files which might be used during execution into the enclave. Thus it doesn't need to interact with the host. Since Python is an interpreted language, the interpreter library (RPython) will handle everything for the source code.  \looseness=-1

\vspace{2pt}\noindent$\bullet$\textit{ EGo}.
EGo~\cite{ego_offical} is a framework for building confidential apps in Go. EGo simplifies enclave development by providing two user-friendly tools:
ego-go, an adapted Go compiler that builds enclave-compatible executables from a given Go project - while providing the same CLI as the original Go compiler; ego, a CLI tool that handles all enclave-related tasks such as signing and enclave creation.


\vspace{2pt}\noindent$\bullet$\textit{ GOTEE}.
GOTEE~\cite{ghosn2019secured} extends the Go language to allow a programmer to execute a goroutine within an enclave, to use low-overhead channels to communicate between the trusted and untrusted environments, and to rely on a compiler to automatically extract the secure code and data. It uses its own security runtime to provide the syscall interpositions.

\vspace{2pt}\noindent$\bullet$\textit{ AccTEE}.
AccTee~\cite{goltzsche2019acctee} embeds a home-baked WASM interpreter written in JavaScript in the enclave, which serves as a sandbox for resource accounting. The JavaScript code is supported by V8 engine\cite{v8_official}, which is built upon SGX-LKL. 

\vspace{2pt}\noindent$\bullet$\textit{ TWINE}.
TWINE~\cite{menetrey2021twine} leverages WebAssembly-Micro-Runtime (WAMR) ~\cite{WAMR} to run WASM code with WASI support. C/C++ and Rust source code developed for Linux can be easily compiled into WASI target, and TWINE also provides a SQLite example.

\vspace{2pt}\noindent$\bullet$\textit{ Enarx}.
Enarx~\cite{enarx_official} is another WASI-compatible WASM runtime on TEEs. Its runtime is powered by wasmtime~\cite{wasmtime_official} to back the trusted application running in the sandbox. Enarx is designed for multiple TEEs and works on AMD SEV and Intel SGX. \looseness=-1




We summarize these TEE containers in Table~\ref{tab:middleware-category}.

\subsection{Threat Models}

All Tcons are designed under one of the following two threat models, either trusting the application it hosts or not. 

\vspace{3pt}\noindent\textbf{Trusted application model (TAM)}. The design of SGX is based upon the assumption that the code running inside an enclave is trusted and that the outside is untrusted~\cite{costan2016intel}. Also physical attacks on RAM and CPU are included in the threat model, whereas denial-of-Service (DoS) attacks are out of scope. Graphene-SGX, SCONE, Occlum's current version, SGX-LKL, and Ratel all follow this traditional threat model.

\vspace{3pt}\noindent\textbf{Untrusted application model (UAM)}. SGX does \textit{not} protect enclave data from untrusted enclave code but many application scenarios demand accommodation of untrusted programs: for example, a healthcare information service that concurrently runs multiple clients' query code within an enclave~\cite{ahmad2021chancel} is expected to confine such code and isolate it from other programs. So Tcons like Ryoan and its sequence Chiron~\cite{hunt2018chiron}, Occlum's original version, Deflection and Chancel are designed to isolate the code from different users and from the sensitive data it is not authorized to access. 
Among the Tcons mentioned above, some of them (Occlum and Chancel) claim that they can protect the computing environment for concurrent use of services by different users.


Most existing Tcon designs claim side channels outside their threat models. 


\subsection{Ocall/Ecall/Exception Interfaces}
\label{susec:external-interface}

To ensure isolated execution, a Tcon is expected to secure its interfaces for its hosted application to communicate with the untrusted OS.  Such interfaces serve the purposes like syscalls, context switching and resource management (e.g., memory management, I/O control and exception handling).  
These interfaces fall into two categories: those facing the application (called \textit{app-middleware interface} or \textit{AMI}) and those facing the OS (called \textit{middleware-host interface} or \textit{MHI}). MHI includes \textit{Ocall} (Outside Call) referring that the caller in enclave invokes a function outside, \textit{Ecall} (Enclave Call) meaning that the caller transfers control flow from outside to the enclave, and exception which may lead to  Asynchronous Exit Events(AEX). MHIs are mainly implemented using Ocalls.


\begin{figure}[]
	\centering
	\setlength{\abovecaptionskip}{0.cm}
	\includegraphics[width=.32\textwidth]{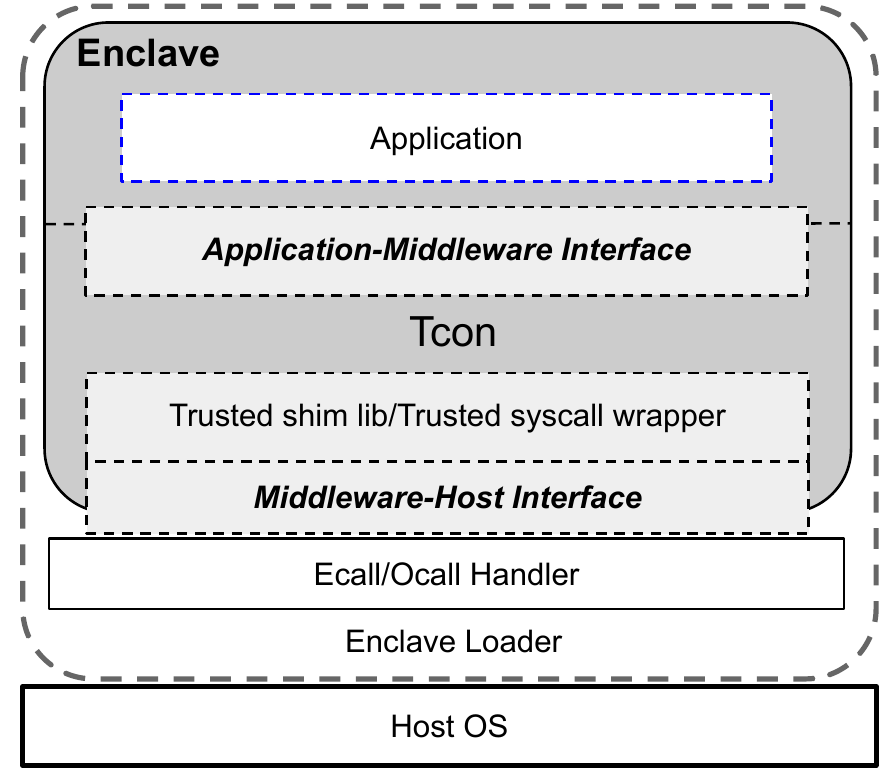}
	\caption{Interface Design of Most Tcons}
	\label{fig:interface-design}
 	\vspace{-15pt}
\end{figure}

\vspace{3pt}\noindent\textbf{Application-middleware interface}.
To mediate binary code's interactions with the OS, a TEE container wraps its LibOS or libc to manage the calls issued by the code. This control can also be done on the WebAssembly System Interface.

\vspace{2pt}\noindent$\bullet$\textit{ LibOS wrapper}. The LibOSes in Occlum~\cite{shen2020occlum}, Graphene-SGX~\cite{tsai2017graphene} and SGX-LKL~\cite{purdila2010lkl} are wrapped to protect their hosting applications.
Some of them implement in the user space a set of kernel-mode functionalities in traditional OSes. Examples include Drawbridge~\cite{porter2011rethinking} used in Haven~\cite{baumann2015shielding}, the Linux kernel library (LKL)~\cite{purdila2010lkl} for SGX-LKL, and  Graphene~\cite{tsai2014cooperation} for Graphene-SGX. 
Most AEXs caused by in-enclave applications are handled by the LibOSes. \looseness=-1

\vspace{2pt}\noindent$\bullet$\textit{ Libc wrapper}. Tcons like SCONE, Chancel, Deflection, and Ryoan utilize the shim libc layer to control the functions an in-enclave application can invoke. Specifically, these Tcons compile the user's source code and when linking it to the libc, they inject the wrapper to manage the individual functions the code is allowed to use.   
The libc wrapper can further request the untrusted OS to complete related functionalities through Ocall, transferring control to the outside of the enclave for executing a syscall.

\vspace{2pt}\noindent$\bullet$\textit{ WASI}. The WebAssembly System Interface (WASI)~\cite{wasi_official} provides OS-like features (e.g., filesystems, sockets) to WebAssembly (WASM)~\cite{wasm_official} bytecode. Tcons running WASM applications (AccTee, TWINE and Enarx) use WASI as AMI. 

\vspace{3pt}\noindent\textbf{Middleware-host interface.~\footnote{Panoply~\cite{shinde2017panoply} and BesFS~\cite{shinde2020besfs} use another kind of MHI - shim POSIX wrapper. The wrapper itself can be linked to the target application, and it forwards the syscall to the outside untrusted libc wrapper. However, Panoply and BesFS act more like SDKs than containers. So, we do not include them in our survey.}}
A Tcon relies on the untrusted OS kernel to perform the tasks the user-land enclave cannot handle, such as I/O and memory management. For this purpose, nearly all  Tcons utilize Ocall stubs to request services from their host kernels, though the specific Ocalls exposed and syscalls supported can vary across different Tcons. Particularly, a Tcon can handle some syscalls inside its enclave (e.g. LibOS can process \texttt{getpid}) and use some syscalls to replace other calls with similar functionalities. In this way, the Tcon can ``squash'' or ``distort'' a syscall when it comes out of the enclave, but still ensure an application's proper execution. 

\vspace{2pt}\noindent$\bullet$\textit{ Ocall stub}.
The Ocall stub is an interface inside the enclave for invoking a syscall and other privileged instructions outside the enclave.
As shown in Figure~\ref{fig:interface-design}, a Tcon using libc (SCONE, Chancel and Deflection) forwards syscalls outside the enclave through an \textit{trusted syscall wrapper}, while the Tcon running LibOS (Graphene-SGX, Occlum and SGX-LKL) sends syscalls to a \textit{trusted shim lib}. Both the wrapper and the shim lib are  a thin software layer that handles privileged operations, that contains a set of Ocall stubs.   
Such a stub delegates an inside request (such as a syscall) to an outside entry/exit Ocall handler library. Most Tcons use Ocall stubs as the MHI, with Ryoan being the only exception we are aware of.


\vspace{2pt}\noindent$\bullet$\textit{ DBT rewriting and forwarding}. This is the unique MHI of Ratel. Ratel can intercept all entry/exit points and simulate context switches and further runs a Dynamic Binary Translation (DBT) engine to 
update binary code on-the-fly by rewriting instructions (e.g., converting a syscall instruction to a stub or library function call) before forwarding a call outside the enclave when necessary. 



\subsection{Internal Isolation}


Under the UAM, the application hosted by a Tcon is untrusted. Therefore, the container needs to enforce a logical separation between the code and the data with different levels of trust. For this purpose, existing Tcons either implement SFI or use WASM-based sandboxing.  





\vspace{3pt}\noindent\textbf{Software-based Fault Isolation}.
Software-based Fault Isolation is a standard solution to in-enclave isolation. For example in Ryoan~\cite{hunt2016ryoan}, memory reads/writes, and control-transfer instructions are all confined in an NaCl based container. 
At a high level, SFI enforcement checks every dangerous instruction (such as load and store) to ensure its safe use. Its implementation on binary code is typically through either Dynamic Binary Translation (DBT) or Inlined Reference Monitors (IRM)~\cite{tan2017principles}.
DBT uses an efficient interpreter to interpret instructions in the target program. For each instruction, the interpreter checks that it is safe according to some policy before the instruction is executed.
On the other hand,
IRM uses a static rewriter to instrument a program with inlined security checks. When the instrumented program executes, the checks before dangerous instructions prevent any policy violation. 
Occlum, Chancel, Deflection, and Ryoan all host a lightweight IRM verifier in the enclave. Occlum, Chancel and Deflection~\cite{liu2021practical} also implement  SFI in their compilers.
To reduce the TCB as much as possible, they do not trust the compiler tool-chain, but rely on the in-enclave verifier to enforce SFI policies.

\vspace{3pt}\noindent\textbf{Sandboxes using WebAssembly}.
WebAssembly~\cite{wasm_official} is an emerging standard defining the binary format for a stack-based virtual machine. It was first adopted in web and later supported as a compilation target of LLVM. WASM code runs in a sandboxed environment with some important security features~\cite{wasm_security}, which can be leveraged by Tcons for separating untusted code from each other and from sensitive data outside its privilege. Such Tcons usually compile source code to a webassembly target, load the WASM module into a sandbox and use an interpreter or a JIT/AoT compiler to execute target. Such containers include AccTee, TWINE, and Enarx.

\subsection{Other Security Properties}

Other security properties expected from a TEE container include side-channel control, attestation, remote attestation (RA) in particular, and secure storage~\cite{intel_pfs}.    


\vspace{3pt}\noindent\textbf{Side channel protection}. Side channel is not considered in the original design of SGX~\cite{hoekstra2013using}, which however has become an important security challenge for TEE-based applications. Most prominent threats to TEE include those on the OS layer, such as page-based attacks~\cite{xu2015controlled} and interrupt-based attacks~\cite{he2018sgxlinger}, and those on the micro-architectural level, such as various cache-based attacks~\cite{schwarz2017malware,gotzfried2017cache,moghimi2017cachezoom}, exploits on speculative execution~\cite{Spectre2018}, microarchitectural data sampling~\cite{schwarz2019zombieload}, etc.  Although the latter is better fixed by hardware manufacturers, the developers of Tcons are at the position to put some protection in place against the OS-level threats. However, most of them assume away side channels in their threat models, with Graphene stealthily implementing some mitigation, as found in our research.


\vspace{3pt}\noindent\textbf{Attestation}.
Attestation is the key part of a Tcon. It is the trust foundation of using Tcon remotely, and it should be designed with an Ecall to call the trusted hardware attestation primitives inside the enclave. 
Some Tcons are well designed in this regard. 
For example in Ryoan, before passing sensitive data to Ryoan a user will request an attestation from SGX and verify that the identity is correct.
In many cases, the user also wants secret provisioning to transparently transfer secret keys and other sensitive data to the remote TEE. However, some Tcons (e.g., Deflection and Chancel) have poor implementation on the secret provisioning part. Worse, Ratel has no attestation support. \looseness=-1



\vspace{3pt}\noindent\textbf{Secure storage}.
Confidentiality and integrity of user data is critical and what users pay most attention to. Tcon can leverage TEE data sealing capabilities for secure in-memory and persistent storage.
It is worth mentioning that different Tcons support secure storage differently.
Some of them (e.g., Occlum, SGX-LKL) have a protected file system, while Tcons like Graphene-SGX, SCONE have an set of APIs for encrypt and decrypt file I/O transparently. 
Some TEE containers such as Ratel, Deflection, etc. do not have the support of secure storage. \looseness=-1

\section{TEE Container Analysis: Motivation and Methodology}
\label{sec:motivation}

\subsection{Needs for In-depth Analysis on TEE Containers}

The literature of existing Tcons only gives an incomplete, often coarse-grained picture of protection implemented by these containers, which is far from enough to understand their security properties.  Particularly, it is less clear whether indeed these Tcons can ensure isolated execution -- the fundamental security requirement for any TEE design.  Following we present the missing links in the public description of these containers' security designs, highlighting the research questions that motivated our experimental analysis on them.   


\vspace{3pt}\noindent\textbf{Deficits in understanding}. As discovered in our research, Tcon documents (papers, developer manuals, and others) tend to miss some important aspects of a container's security protection, and for those they cover,  often technical details are missing, raising the question whether the Tcon has been correctly implemented. Particularly, for isolated execution, little information has been given on how individual Tcons manage syscalls: for those running trusted applications, whether the return value from the untrusted OS has been properly checked to prevent the exploits like the Iago attack; for those hosting untrusted applications, whether their syscall parameters have been inspected and sanitized to detect or defeat a covert-channel attack. Another problem is Ecall interface, whose security protection is often not detailed by Tcon publications. 

Also concerning are these Tcons' use of SFI for internal isolation. Although SFI is known for its runtime efficiency and strong guarantee in enforcing data-access and control-flow policies, it is often hard to do right. Most published documents of today's Tcons do not offer detailed account on how their SFI implementations work. For example, it is less clear whether all branching instructions are fully mediated by these Tcons, which could have a significant performance impact. As another example, the publications of some Tcons do not mention how some critical instructions like ENCLU are controlled, which if unprotected, can be used as a gadget in an exploit~\cite{biondo2018guard}.  Also, none of the prior work reports whether libaries uploaded at runtime have been properly instrumented and controlled. 

Further important to isolated execution is side-channel control. Although all existing Tcons assume away this security risk in their threat models,  still we want to understand whether the presence of the Tcon middleware could make an OS-level side-channel attack harder to succeed.

\vspace{3pt}\noindent\textbf{Research questions}. In our research, we aimed at demystifying the isolation protection implemented by existing Tcons, given the central role this property plays in the TEE's security assurance. More specifically, we intended to answer the following questions through an experimental analysis: 

\begin{packedenumerate}
    \item[RQ1.] What Ecall/Ocall interfaces have been implemented in existing TEE containers? Are they well-protected?
    \item[RQ2.] How effective is the internal isolation implemented by existing Tcons? Does the protection fully cover in-enclave attack surfaces?
    \item[RQ3.] Have existing Tcons raised the bar to OS-level side-channel attacks? 
\end{packedenumerate}

Due to the lack of public information, these questions can only be answered by an experimental analysis on these Tcon implementations, as elaborated below. 


\subsection{Overview of Our Study}

Here we present the methodology of our experimental analysis and the settings of the study. 

\vspace{3pt}\noindent\textbf{Methodology}. To answer the research questions identified, we designed a methodology that utilizes both automated analysis and manual validation. 
More specifically, to understand the interface protection of existing Tcons (RQ1), we developed an automated analyzer to fuzz the syscall interfaces implemented in different Tcons, and further reviewed the source code of other Ecall/Ocall interfaces. To answer RQ2, we utilized the analyzer to test these Tcons' SFI implementations based upon a set of security policies expected to be enforced for internal isolation~\cite{sinha2016design}. To find out whether Tcons raise the bar to an OS-level side-channel attack (RQ3), our analyzer runs a set of benchmarks built on known attacks against these Tcons.


\vspace{2pt}\noindent$\bullet$\textit{ Automatic in-depth analysis}. At the center of our methodology is a \textit{2-piece} analyzer, including the components both inside and outside a Tcon.  So a test input can be injected from the application hosted by the Tcon or from the OS kernel and received at the other end to evaluate the interface protection of the container (e.g., sanitizing parameters or return values). 


\vspace{2pt}\noindent$\bullet$\textit{ Manual validation}. Some security-critical designs vary significantly across containers, making it hard to do an automated analysis. This typically happens to Ecall interfaces, which are meant to upload different content (configuration, attestation data, code) into the enclave. For instance, to configure a Tcon, the user of Graphene-SGX needs to write a manifest file that specifies what and how unmodified binaries and libraries are loaded into the enclave; for Occlum and SGX-LKL, one is expected to prepare a disk image for the application, which will be imported into the Tcon.  In our research, we looked into some easy-to-inspect features of a Tcon's Ecall/Ocall interfaces, whose source code is typically organized in a similar way for the same type of containers (LibOS or libc-based). For example, LibOS-based Tcons usually include the code of their Ecall/Ocall interfaces under the directory `PAL' (Platform Abstraction Layer) or `interface'.


We focused in our research on two key Ecall interfaces -- remote attestation and code loading, and reviewed their source code for each Tcon studied. In the meantime, to complement our analyzer, we also manually inspected some Ocall-related code for the feature hard to evaluate automatically. For example, we found the total number of the syscalls each Tcon supports and those exposed to the OS. We also checked whether the Ocall interface of each Tcon supports running of raw syscall instructions.  


Finally, we contacted authors of Tcon papers and developers of selected containers to get their feedback on our findings. This helps validate the discoveries made in our research and identify  key takeaways for future development of Tcons. 




\vspace{3pt}\noindent\textbf{Tcon collection, install and experiment settings}. In our research, we used our methodology to evaluate 8 Tcons, including Graphene-SGX, SCONE, Occlum, SGX-LKL, Chancel, Deflection, Ratel and Ryoan. We selected those containers since our analyzer is developed using Rust and for operating under Linux, which are supported by these Tcons. The source code of these Tcons were collected from Github (including two versions of Occlum, see Section~\ref{subsec:existingtcons}), with some exceptions. First, we only requested SCONE's community version, since its commercial implementation is not publicly available. Second, we contacted the developers of Chancel and Ryoan for their code. The Ryoan team released partial source code at Github~\cite{ryoan_code}, which however did not work. So we had to analyze Ryoan manually. The developers of Chancel gave our permission to access their private repository. 

For our experiment, we set up Tcons based upon their installation guides.  Particularly, we  utilized a machine that supports both SGX and MPX to run Occlum's artifact evaluation version, and a system with 64GB memory to evaluate Deflection that needs a large memory~\cite{deflection_code}. Also Graphene-SGX and Occlum require an FSGSBASE kernel patch, so we installed them on Linux kernel 5.9. 

\section{TECUZZER: Fuzzing TEE Containers}
\label{sec:fuzzer}

In this section, we elaborate the design and implementation of TECUZZER, our Tcon analysis tool.

\subsection{Design and Implementation}
\label{subsec:fuzzer}


\noindent\textbf{Framework design}. 
TECUZZER is a two-piece TEE container analyzer, with a user-level component (U-part) and a kernel-level component (K-part). 
The U-part contains a code generator that creates test cases; the K-part catches the test requests penetrating the Tcon for inspection, and then sends back crafted return values to the container when necessary.  

To intercept the test requests delivered to the kernel and control the return value, TECUZZER utilizes \textit{strace}~\cite{strace} as the first stage interceptor and \textit{ftrace} as the second stage interceptor, which intercepts in-kernel syscall/fault processing. Specifically, the ftrace framework~\cite{rostedt2014ftrace} is not only for monitoring each syscall's service routine (e.g., \verb|sys_read| for syscall \verb|read|) for call parameters, but also for hooking some system functions (such as \verb|__do_page_fault|) for faults and interrupts. 
These two components communicate with each other through a communication channel built on  \textit{debugfs}, a RAM-based file system specially designed for transferring information between the kernel space and the user space. 
Further our design uses a Linux built-in logger as the recorder to collect data for analysis.

%


Below we explain three tasks performed in our study: syscall fuzzing, SFI functionality checking, and side channel analysis. 

\begin{figure}[]
	\centering
	\setlength{\abovecaptionskip}{0.cm}
	\includegraphics[width=.45\textwidth]{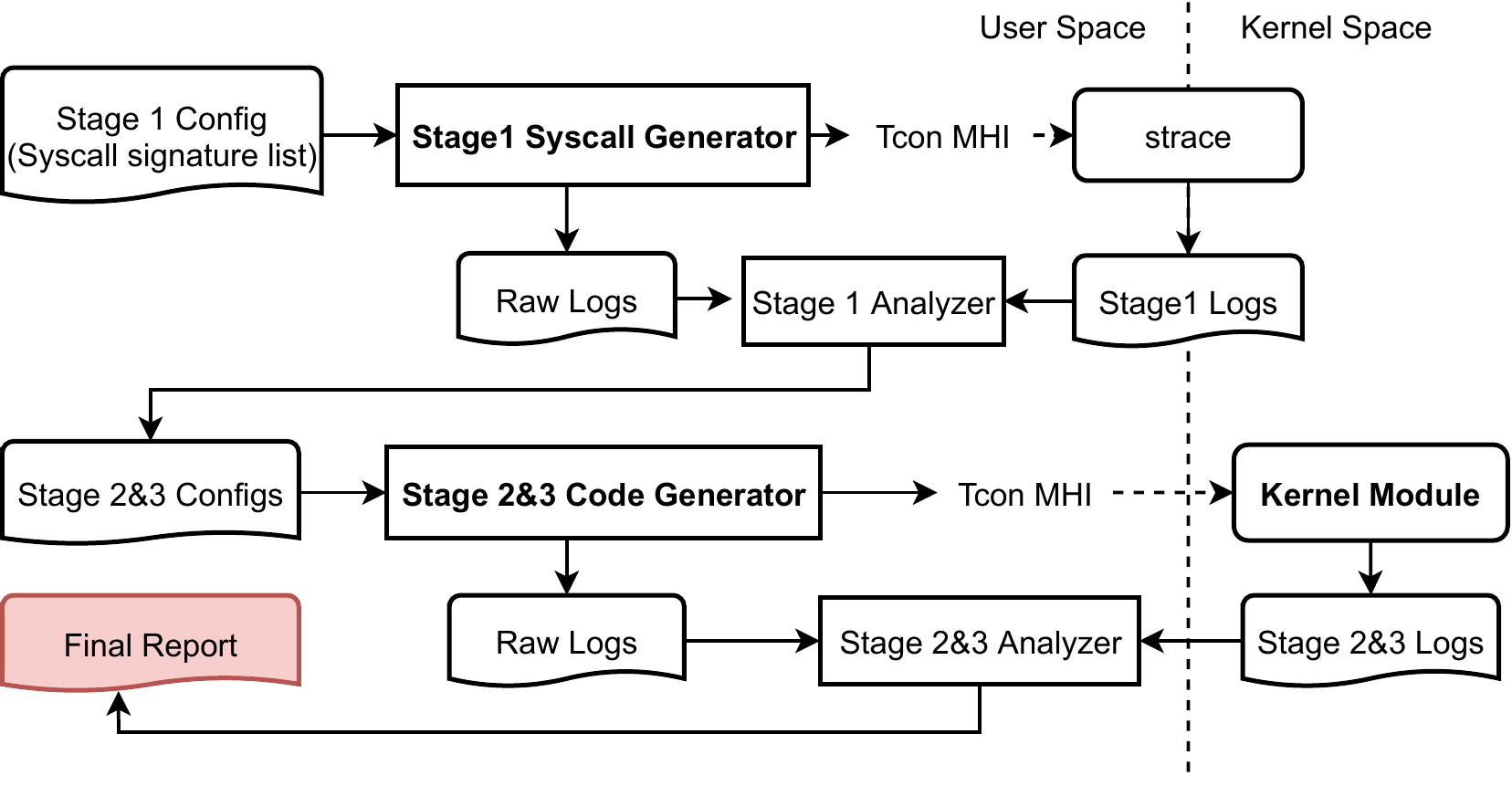}
	\caption{Workflow of Syscall Fuzzing} 
	\label{fig:fuzzer-pipeline}
 	\vspace{-15pt}
\end{figure}

\vspace{3pt}\noindent\textbf{Syscall fuzzing}.
Syscall fuzzing tests for each syscall, how much protection, if any, is enforced by the Tcon.
Unlike prior tools such as Trinity~\cite{trinity_repo} and Syzkaller~\cite{syzkaller_repo} which are designed to fuzz Linux kernel, TECUZZER differs in a few important ways. 
First, TECUZZER's target is container runtime layer.
Second, TECUZZER follows a unique two-piece design to perform both U-part to T-part and T-part to U-part testing. Third, it follows a multi-staged and feedback-guided design to improve efficiency. Third, it is written in Rust, so it can ensure the consistency in the types of K-part and U-part variables during fuzzing. The workflow of the syscall fuzzing component is shown in Figure~\ref{fig:fuzzer-pipeline}.

\vspace{2pt}\noindent$\bullet$\textit{ Stage 1: Which syscalls are served by untrusted OS}.
At Stage 1, we check if a syscall is handled within the Tcon or by the untrusted OS. Some syscalls are never processed by the untrusted OS: they are totally served in the Tcons or the support for the syscalls is missing. Others may be served in the Tcons at specific circumstances or may be forwarded to the OS, either with or without distortion. For example, a LibOS can convert \verb|read| to \verb|pread| by adding an addition parameter, \textit{offset}. Besides, the validity and correctness checks (e.g. invalid pointer check in Graphene-SGX and Occlum) is also an obstacle. \looseness=-1

Stage 1 uses the function signature of each syscall, as documented in the Linux manual~\cite{linux_mannual2}, to generate initial test cases for each syscall. Then, we utilize strace~\cite{strace} to record traces of syscalls that are exposed to the untrusted OS. One technical challenge here is to generate \textit{semantically correct} syscalls, which can be handled by the OS with no error returned. TECUZZER encodes the syscall signatures and all definitions of structs used in it with methods to efficiently generates them from random number generator with seed. \looseness=-1

The result of Stage 1, a list of exposed syscalls (including both not served and partially served syscalls) is passed to Stages 2 and 3. Note that the multi-stage design improves efficiency: Stages 2 and 3 only need to focus on the exposed syscalls. \looseness=-1



 


\vspace{2pt}\noindent$\bullet$\textit{ Stage 2: Which parameters are sanitized}.
At Stage 2, the analyzer checks which parameters can leak information in syscalls and what sanitization mechanisms are applied by the Tcons. To do so, 
TECUZZER compares the parameters from both the U-part and K-part (intercepted by the kernel hooking/logging module).
TECUZZER uses 3 different fuzzing strategies in Stage 2.
First, it tests semantically correct syscalls.
Second, it expands the search by considering all values that match the corresponding types in signature. For example, \textit{enum} is treated as random \textit{int} and the length of a buffer can be erroneous. Third, it generates all-random parameters. During fuzzing, TECUZZER logs syscalls parameters and related data structures sent by U-part and received by K-part. The data structures are usually passed to the kernel as pointers and may contain pointers to other data structures (nested structs), whereas our logger dereferences these pointers recursively and record data structures comprehensively. The logs are fed to the analyzer to resolve protections on syscall parameters and which fields in parameters can leak information for the tested syscalls. \looseness=-1

\vspace{2pt}\noindent$\bullet$\textit{ Stage 3: What a malicious kernel can do?}
At Stage 3, the analyzer calculates the bandwidth of leaking information through syscalls and check if there is any sanitization in the Tcons for return values. The workflow of Stage 3 is identical to Stage 2, but the K-part acts more actively by modifying return values. This is required both to construct effective covert channels, as well as to check sanitization on return values. For example, the K-part can skip serving \verb|nanosleep| and return \textit{0} directly when receive from Tcon, and meanwhile modify the \textit{rem} \textit{timespec} struct returned to the Tcon. So, we can construct an efficient covert channel and forge return values, including which is pointed by syscall parameter, at the same time.
%
%
%
On the one hand, to compute information leakage, Stage 3 analyzer calculates the covert channel bandwith by multiplying leaked bits in each call by the syscall speed rate. On the other hand, it also checks if and how the Tcons validate and modify return values by comparing logs of return values from K-part and U-part. \looseness=-1
\noindent\textbf{Implementation}.
To avoid possible pitfalls on types and human errors, we implement the fuzzer in Rust. Thanks to Rust's powerful macro system and \textit{trait} feature, the Rust implementation allows us to avoid manually writing redundant syscall wrapper and generation function for a syscall. We manually implement the random generation methods for structs used in syscalls, and the macro helps us to write code calling these methods to generate all parameters for each syscall as we encode its signature. The procedural macro can implement relatively simple but error-prone traits automatically. Also, the structs with their own constructor and destructor methods can be reused across different syscalls, which helps the developer save engineering effort greatly and benefits secondary development. \looseness=-1

TECUZZER tests stateful syscalls in stateless fashion. Many syscalls are inherently stateful, depending on the success of other syscall(s). For example, almost all file operation-related syscalls require a file descriptor (\verb|fd|) as an argument, which is returned from \verb|open| when succeed, and memory-related syscalls need to operate on a specific memory region, which may corrupt the fuzzer itself if it's randomly choosen. TECUZZER opens valid  (\verb|fd|), allocates new memory regions, (e.g. \verb|mprotect|) and spawn new threads for fuzzing file-system, memory-management, and thread-related syscalls respectively to prepare ``clean'' testing environments. 
Moreover, for sanity and avoiding interfering the following fuzzing cycles, the fuzzer cleans the footprint at the end of each cycle, including \verb|munmap| the memory, \verb|close| the \verb|fd| , and join the thread. \looseness=-1





\vspace{3pt}\noindent\textbf{SFI functionality checking}. To analyze the SFI implementation of various Tcons with SFI support (including Occlum~\cite{shen2020occlum}, Chancel~\cite{ahmad2021chancel} and Deflection~\cite{liu2021practical}), TECUZZER uses a carefully crafted test suite of SFI functionality (more details covered in Section~\ref{sec:sfiflaws}). The test suite covers important components of SFI, including security checks on memory store, \texttt{RSP} spill, direct/indirect branching, \texttt{ENCLU} gadget, and instrumentation of necessary libraries. The test suite contains four Proof-of-Concept (PoC) attack functions designed specifically to break these components. Specifically, these PoC functions try to write to an unauthorized memory location, to jump into the middle of instrumentations, to execute an \texttt{ENCLU} instruction, and to insert a dangerous code (e.g., writing to the outside of the enclave) in the musl-libc which used in those SFI-based Tcons. If any of these attempts succeed, TECUZZER reports a vulnerability in SFI implementation.

\vspace{3pt}\noindent\textbf{Side channel analysis}.
To understand side channel leaks through Tcons, we built two test cases both into the U-part and K-part of TECUZZER. These two test cases include the page-fault attack~\cite{xu2015controlled} and other exception-based attacks~\cite{van2017sgx}. We implemented both attacks on PTEditor~\cite{pteditor} and SGX-STEP~\cite{van2017sgx}
to help handle the page fault and interrupts.






\subsection{Discussion}

\vspace{3pt}\noindent\textbf{Supporting more Tcons}.
The current version of TECUZZER can fuzz 7 Tcons, including Graphene-SGX, SCONE, Occlum, SGX-LKL, Chancel, Deflection, and Ratel. Since TECUZZER generates test cases written in Rust, we can measure Tcons which either support unmodified binaries or compile Rust source code, such as Graphene-SGX, SCONE, Ratel, etc.
As for Tcons that only support C (Occlum AE, Chancel, Deflection, etc.), we first compile the U-part code (Rust) to an IR-level code, and then compile the IR code to a relocatable file using Tcon's target-level LLVM pass. Finally the relocatable file can be ported into Chancel/Deflection's enclave loader.
Currently, TECUZZER does not support Tcons using Go or WASM as an input; this can be our future work.

\vspace{3pt}\noindent\textbf{Supporting more syscalls}.
While the fuzzer has the ability to discover what protection a Tcon's syscall interface has, it would be desirable to have more extensions. Currently we can fuzz 45 syscalls in total. We choose the most important 35 ones according to the survey~\cite{tsai2016study} plus 10 random ones. The number can be more when  investing more manpower and/or accepting community support.

\vspace{3pt}\noindent\textbf{Extending to VM-based TEE}.
Since running an application on virtual machine-based TEE still relies on host/hypervisor intervention, information disclosure can happen explicitly or covertly through the interactive interfaces. To that end,  a analysis tool to measure whether the future TEE container can be trusted will be necessary.
Although our TECUZZER cannot be copied onto the virtual machine-based TEE directly, the idea of our solution is still applicable since the problems are similar. Developing new detection tools on the hypervisor of a upcoming TEE is the future work.

\section{Security Analysis and Findings}
\label{sec:findings}

\begin{figure*}[]
	\centering
	\setlength{\abovecaptionskip}{0.cm}
	\includegraphics[width=.8\textwidth]{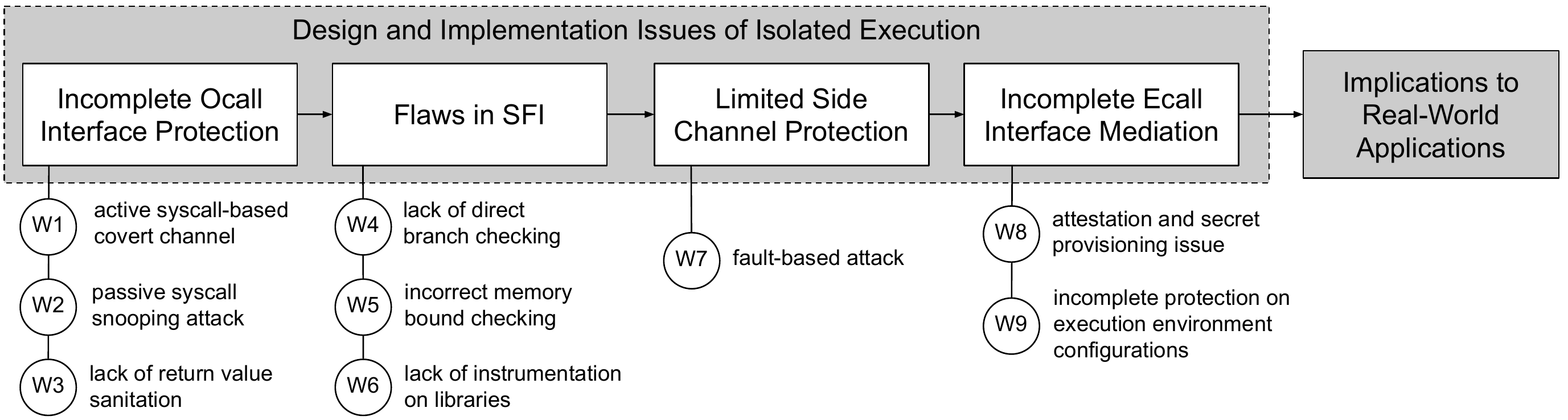}
	\caption{Common Pitfalls of Tcons} 
	\label{fig:common-pitfalls}
 	\vspace{-15pt}
\end{figure*}


In our study, we applied our methodology to all 8 Tcons, running TECUZZER on 7 of them (except Ryoan, whose Github executable cannot run). Our manual analysis took 2 cybersecurity professionals 3 months to accomplish. Also we spent 14 days on configuring the environments and running TECUZZER on the 7 Tcons. In the end, our analysis reports 9 weaknesses in these popular Tcons. 
Among them,  most (Weakness 1, 3, 4, 5, 6, 7) were identified automatically. 
Although some of these problems may seem obvious at first glance, they are quite pervasive among all Tcons, indicating more fundamental causes behind.  We organize our findings in Figure~\ref{fig:common-pitfalls} and elaborate on them as following.



\subsection{Incomplete Ocall Interface Protection}

\begin{table}[]
\belowcaptionskip=-10pt
\caption{Syscall Interfaces Statistics}
\label{tab:syscall-interfaces}
\begin{center}

\begin{tabular}{|p{2cm}|p{1.7cm}|p{1.7cm}|p{1.7cm}|}
\hline
Tcons measured by 2021.06  & \# of supported syscalls & \# of exposed OCalls & Native syscall support \\ \hline
Graphene-SGX & 161                      & 42                   & $\checkmark$                              \\ \hline
SGX-LKL      & 267                      & 7                    & $\times$                               \\ \hline
Ratel        & 212                      & 193                  & $\checkmark$                               \\ \hline
Occlum       & 120 + 5*                & 37                   & $\checkmark$                            \\ \hline
Deflection   & 36                       & 36                   & $\times$                               \\ \hline
Chancel      & 38                       & 38                   & $\times$                               \\ \hline
Ryoan        & 279                      & 279             & $\times$                              \\ \hline
\end{tabular}
\end{center}

* Occlum-specific system calls

\vspace{-15pt}
\end{table}

We summarize in Table~\ref{tab:syscall-interfaces} the syscall interfaces controlled by different Tcons as detected in our study. 
As we can see from the table, libc-based Tcons (Deflection, Chancel and Ryoan) simply forward all syscalls to the kernel while LibOS-based containers (Graphene-SGX, SGX-LKL, and Occulum) handle a large number of syscalls within the middleware, so their exposed MHIs are much narrower: e.g., out of the 161 syscalls Graphene supports, only 42 are handed over to the kernel.  

\vspace{3pt}\noindent\textbf{Weakness 1 - active syscall-based covert channel}. For the Tcon under the threat model of UAM, where its host application is untrusted, covert channel is a risk that cannot be ignored. Through the channel, the application could exfiltrate in-enclave data to the untrusted OS. This could be done using syscall parameters. Although these containers are supposed to offer some level of protection, as claimed in their papers~\cite{ahmad2021chancel, hunt2018ryoan, liu2021practical}, our analysis shows that often there is no defense. \looseness=-1


Chancel and Deflection neither check nor change any call parameter; they just simply forward each syscall to the host OS. The authors of the Ryoan paper claim that the container does not allow the \verb|read| syscall from an untrusted application to be directly sent to the kernel, since the pointer the call passes to the OS could reveal a large chunk of in-enclave data. However, our analysis did not find any such protection in its Github code. \looseness=-1 

\vspace{2pt}\noindent$\bullet$\textit{ Impact Analysis}.
We ran TECUZZER to evaluate the protection implemented by these Tcons on the syscall surface and also measured the bandwidth of information leaks through call parameters (Section~\ref{subsec:fuzzer}). Only Occlum is found to have certain protection. Parameters like \textit{pollfds} and \textit{timeout} are checked in syscall \verb|poll|. And \textit{msg\_flags} is sanitized in \verb|sendmsg| and \verb|recvmsg|.

It turns out that almost all exposed syscalls can be used to exfiltrate sensitive data, with \verb|futex| and \verb|nanosleep| being the ones with the largest bandwidth (3Mbps), since the malicious kernel can complete those syscalls instantly, without returning any useful values. 


\vspace{2pt}\noindent$\bullet$\textit{ Mitigation}.
Tcons should only allow necessary syscalls (e.g., network I/O and file I/O), together with their wrappers for security control (e.g., applying encryption and limiting the range of syscall arguments). Specially, the wrapper for \texttt{send} encrypts the message to be delivered and pads it to a fixed length. Further, the wrapper can put a constraint on the length of the result to control the amount of information disclosed to the code provider: e.g., only 8 bits can be sent out.


\vspace{3pt}\noindent\textbf{Weakness 2 - passive syscall snooping attack}. Most Tcons cannot handle I/O requests, especially disk I/O. So they have to resort to the untrusted kernel for serving the requests. This however exposes a lot of information to the OS. Using file operations as an example, the kernel knows which file is open and being processed and even the offset at which a read or write operation takes place. This could cause information leaks even from an encrypted file system, since sensitive data could still be learnt from individual files' meta-data and the pattern of the accesses they receive. Although this problem has already been known~\cite{ahmad2018obliviate},  less clear is whether any protection has been implemented in today's Tcons to mitigate its risk.    


\vspace{2pt}\noindent$\bullet$\textit{ Impact analysis}. By analyzing these Tcons' code, we are surprised to find that indeed some Tcons include some protection against the threat, even though none of their papers have mentioned it. Specifically, Occlum and SGX-LKL each runs an in-enclave file system (FS), so the operations on the files are not exposed to the untrusted OS. Ryoan also seems to be well protected since it uses in-memory POSIX APIs to access preloaded files, which is unobservable to the OS. Graphene-SGX runs a hybrid FS~\cite{ahmad2018obliviate}, where some files are kept inside the enclave while others are outside. As a result, the operations on the external files still expose access patterns through I/O. Unlike such LibOS-based Tcons, Chancel, Deflection, and Ratel do not hide file access patterns at all.

\vspace{2pt}\noindent$\bullet$\textit{ Mitigation}. In-enclave FS seems to be a good solution to the syscall snooping attack.  A more generic alternative could be oblivious RAM (ORAM~\cite{sasy2018zerotrace,ahmad2018obliviate,ahmad2019obfuscuro}), which though much more heavyweight, could protect not only disk I/O operations but also the network operations exposed to the same threat. 


\vspace{3pt}\noindent\textbf{Weakness 3 - lack of return value sanitization}. In addition to the outgoing parameters, a Tcon is also supposed to control the syscall return values, so as to mitigate the risks of exploits such as the Iago attack~\cite{checkoway2013iago}. 
Our analysis through TECUZZER shows that some Tcons leave this attack surface unprotected and therefore are vulnerable to the attacks through the return values.  \looseness=-1


\vspace{2pt}\noindent$\bullet$\textit{ Impact analysis}. Graphene-SGX and SCONE are found to implement security checks well on the return values, while some others do not.
In general, \verb|mmap| and \verb|munmap| are most noticed. Graphene-SGX, SCONE, SGX-LKL, and Ryoan all provide protections on them.
Occlum explicitly claims the Iago attack to be out of its threat model, but it has return value checked in \verb|recvmsg|. Although Chancel mentions that it can defeat FS-related Iago attacks through in-memory FS at runtime, we found that the FS has not been implemented in its released version through manual code review. Deflection and Ratel do not enforce any control on return values. Interestingly, our manual analysis shows that some Tcons actually check the return values of some syscalls \textit{outside} the enclave, by their untrusted components. For example, the return of \verb|write| in Chancel and the return of \verb|getdents| in Graphene-SGX are all inspected in this insecure way.   





\vspace{2pt}\noindent$\bullet$\textit{ Mitigation}. This weakness can be addressed by mandating return value checks for all exposed syscalls within an enclave. A challenge is how to model the legitimate return for each call, which could requires significant manual effort. In the meantime, we found that Ratel, a Tcon not implementing the protection, actually leaves the door open to adding the security check: it provides an interface for each exposed syscall so its user can add protection code there.

\subsection{Flaws in SFI}
\label{sec:sfiflaws}

For the Tcons running under the threat model of UAM, isolation should be enforced even inside the enclave. Our analysis on these Tcons reveals the weaknesses in such protection.  
Note that we also measured \texttt{RSP} spill protection and they all mitigate this appropriately. 
The \texttt{ENCLU} checkings are performed but not elaborated as a weakness, since we believe it is a problem that can be easily fixed.
The checked SFI functionalities and whether these Tcons have respective protections are listed in Table~\ref{tab:sfi}.



\begin{table*}[]
\belowcaptionskip=-10pt
\caption{SFI Functionalities}
\label{tab:sfi}
\begin{center}
\begin{tabular}{|c|c|c|c|c|c|}
\hline
Tcon & Mem store (including DEP) & RSP spill             & Direct branch                & ENCLU                 & Verification on libraries \\ \hline
Occlum AE    & Using Intel MPX *  & $\checkmark$  &  $\checkmark$  & $\checkmark$    & Libc instrumented  \\ \hline
Deflection   & Not complete    & $\checkmark$   &  $\times$     & $\checkmark$     &  $\times$  \\ \hline
Chancel   &  $\times$  & $\checkmark$ &  $\times$  &  $\times$ &  $\times$ \\ \hline
Ryoan **     &  $\checkmark$  & $\checkmark$ & $\checkmark$  & $\times$ &  N/A \\ \hline
\end{tabular}

* Intel MPX is already obsolete and the instruction \texttt{BNDMK} was able to be called maliciously.

** We only evaluate it from its paper and code since Ryoan is not able to run.

\end{center}
\vspace{-15pt}
\end{table*}

\vspace{3pt}\noindent\textbf{Weakness 4 - lack of direct branch checking}. Tcons like Occlum, Deflection and Chancel claim implementation of SFI for in-enclave isolation~\cite{ahmad2021chancel, liu2021practical, shen2020occlum} in a way similar to Proof-Carrying Code~\cite{necula1997proof}: an \textit{untrusted} compiler outside the enclave instruments an application, and a trusted in-enclave verifier checks the instrumentations before running the application to ensure that all its critical operations have been properly guarded. For this purpose, all branching instructions need to be controlled so the program will not jump to the location between the instrumented code and the critical instruction, bypassing the protection. In our research, we did not find the SFI implementation in Chancel. Most interestingly, even though both Occlum and Deflection do safeguard indirect jumps in an instrumented application, their verifiers fail to properly check the direct jumps.      
For example, Figure~\ref{fig:SFI_bypass_code} shows a
code snippet where we forged direct jump destination from Line 2 to Line 14, which leads to an unchecked memory write instruction. Both Occlum and Deflection are found to fail to properly check the dangerous memory write at Line 15: they just check the presence of instrumented guard (Line 3 to 13) and ignore the jump target at Line 14.  


\begin{figure}
\begin{center}
\begin{minipage}{0.5\textwidth}
\centering
\begin{tikzpicture}[scale=0.7, font=\sffamily\bfseries\scriptsize, transform shape]
   \tikzset{
     box/.style    = { rectangle,
                       align           = left,
                       font            = \sffamily\footnotesize,
                       text width      = 12cm, 
}
  }
  \node [box](e1) at (0,0)
  {
\begin{lstlisting}[basicstyle=\ttfamily\small,escapeinside={(*@}{@*)}]
	jmp	.LBB0_1                 # direct branch
.LBB0_1:                        # original destination
	pushq	%rbx                # reserve regs for check
	pushq	%rax
	leaq	(%rcx), %rax        # rax <- destination of mov
	movabsq	$0x3FFF..., %rbx    # check upperbound
	cmpq	%rbx, %rax
	ja	.EXIT_LABEL             # exit if check fails
	movabsq	$0x4FFF..., %rbx    # check lowerbound
	cmpq	%rax, %rbx
	jb	.EXIT_LABEL             # exit if check fails
	popq	%rax                # release regs for check
	popq	%rbx
.LBB0_1:                        # forged destination
	movq	%rax, (%rcx)        # memory write
                    ...\end{lstlisting}
}; 
\end{tikzpicture}
\end{minipage}
\end{center}
\abovecaptionskip=-10pt
\caption{Memory write instrumentation bypass}\label{fig:SFI_bypass_code}
\vspace{-15pt}
\end{figure}





\vspace{2pt}\noindent$\bullet$\textit{ Impact analysis}. Once the security check has been circumvented, a malicious in-enclave applications can defeat all access controls to take over the whole enclave and leak out sensitive user data. We have communicated with both the developers of both Occlum AE and Deflection. The Occlum team informed us that they realized this weakness and addressed it in an unreleased version.



\vspace{2pt}\noindent$\bullet$\textit{ Mitigation}. Checking jump targets during verification can be easily implemented.  A problem, however, is the performance implication, given the pervasiveness of direct jump instructions in a program, which needs a more efficient solution to address. 



\vspace{3pt}\noindent\textbf{Weakness 5 - incorrect memory bound checking}. Under UAM, a Tcon should ensure that the untrusted application cannot transmit sensitive data out of the enclave or compromise Tcon's protection. For SGX1, however, this is challenging since it does not support memory privilege change functionalities at runtime. Therefore the current Tcon design performs memory bound checks to ensure that an application can only write to its own data section. However, our analysis shows that these containers fail to properly enforce this protection. 


\vspace{2pt}\noindent$\bullet$\textit{ Impact analysis}.
As shown in Table~\ref{tab:sfi}, memory store checks are not implemented well on Deflection and Chancel. Occlum AE relies on MPX to enforce the bound checks, which is no longer effective after Intel depreciates MPX. Deflection's boundary checks are coarse, only confining an application to write within the whole enclave EPC memroy range. This is insecure since we found that the in-enclave verifier's heap can be overwritten by our PoC program. Chancel simply does not implement any check of memory bounds.

\vspace{2pt}\noindent$\bullet$\textit{ Mitigation}. Ideally, Tcon should enforce fine-grained memory bound checks, as proposed by MPTEE~\cite{zhao2020mptee} (which however relies on depreciated MPX). This requires the loader to have detailed information about an application's memory layout. Also performing more detailed bound checks will inevitably increase the runtime burden, slowing down the execution of the application, in the absence of hardware support.  \looseness=-1


\vspace{3pt}\noindent\textbf{Weakness 6 - lack of instrumentation on libraries}. An uncontrolled malicious library injected into the enclave can completely defeat the Tcon protection. However, none of the Tcons we studied claim in their papers that they instrument the libraries uploaded to the containers. So in our research, we ran TECUZZER on them to find out whether indeed the protection has been implemented.  


\vspace{2pt}\noindent$\bullet$\textit{ Impact analysis}.  Our research shows that Occlum AE instruments libc, libcxx, libunwind etc., and Ryoan uses it own trusted Ryoan-libc.  However, Chancel and Deflection do not instrument linked libraries, including libc and crypto libraries such as mbedtls. The presence of these unprotected libraries are not detected by their verifiers.


\vspace{2pt}\noindent$\bullet$\textit{ Mitigation}. Tcons need to either instrument untrusted libraries or check their integrity at the loading time (which has not been done in today's Tcon implementations).


\vspace{3pt}\noindent\textbf{Feedback from Tcon developers}. 
We communicated with the developers of Occlum about the differences in two versions.
They believe that SFI is hard to implement comprehensively, and acknowledged that there is a great concern that users are not willing to adopt the SFI-protected Occlum (see their paper~\cite{shen2020occlum}), which needs recompile their source code. So after the depreciation of MPX, which increases the overhead for the SFI protection, they moved away from the UAM threat model. 

\subsection{Limited Side Channel Protection}

As mentioned earlier, all Tcons except Deflection claim that side channels are outside their threat models. Nevertheless, we evaluated the effectiveness of known OS-level side-channel attacks on them, to find out whether the presence of Tcons makes the attacks harder to succeed.

\vspace{3pt}\noindent\textbf{Weakness 7 - fault-based side channel attacks}. We study two attacks based on faults: page-fault attacks~\cite{xu2015controlled} (change of flags on page-table entries to induce page fault for observing access pattern), and AC-fault attacks~\cite{van2020microarchitectural} (change of Alignment Check flags to induce intra-cache line secret data access).  \looseness=-1


\vspace{2pt}\noindent$\bullet$\textit{ Impact analysis}. We found that none of Tcons can withstand page-fault attacks. Interestingly, even though the developers of Graphene-SGX assumed away side channel attacks in their paper~\cite{tsai2017graphene}, we found that the Tcon includes mitigation against the AC-fault attack, which renders the attack less effective. In the meantime, all other Tcons are subject to the attack. \looseness=-1


\vspace{2pt}\noindent$\bullet$\textit{ Mitigation}. Defense against these OS-level side channels has been studied in the past 5 years. Proposed solutions include T-SGX~\cite{shih2017t}, address randomization~\cite{seo2017sgx}, oblivious RAM~\cite{sasy2018zerotrace} and HyperRace~\cite{chen2018racing}, etc. However, it is less clear whether such protection can be integrated into Tcons without undermining their performance.  







\subsection{Incomplete Ecall Interface Mediation}

In addition to automated analysis using TECUZZER, we also manually inspected the Ecall interfaces of those Tcons. Following are our findings. 







\vspace{3pt}\noindent\textbf{Weakness 8 - attestation and secret provisioning issue}. Attestation proves to the user that the remotely executing TEE is trusted, which is followed by establishment of a secure communication channel and provision of secrets to the TEE. Our inspection aims at understanding whether this critical step has been properly implemented.


\vspace{2pt}\noindent$\bullet$\textit{ Impact analysis}. Our code review reveals that Ratel has no attestation support at all, and Deflection and Chancel only provide an RA interface without full implementation. Although SCONE does include a complete RA implementation that can be extended to multiple enclaves~\cite{scone_website}, its community version does not support secret provisioning. As a result, it cannot encrypt its file system using the secret key, which causes its secure storage to fail.  \looseness=-1


\vspace{2pt}\noindent$\bullet$\textit{ Mitigation}. Without thorough implementation of attestation, secure channel establishment and secret provisioning, Tcons today cannot provide the verifiable secure service it promises. So before their real-world deployment, these key functionalities should be included.


\vspace{3pt}\noindent\textbf{Weakness 9 - incomplete protection on execution environment configurations}. In addition to the application running inside enclave, its arguments, environment variables and necessary configuration files should also be evaluated to ensure that their integrity is protected and they will not undermine the security protection in the enclave.


\vspace{2pt}\noindent$\bullet$\textit{ Impact analysis}. Our analysis shows that 
Occlum provides the best protection among all Tcons: it runs \textit{initfs} to include all arguments, environment variables and configurations in a remote attestation, to ensure their integrity, and then utilizes \textit{unionfs} as an encrypted secure storage to protect them and further check and store these metadata of the code uploaded at runtime. SGX-LKL also packs the binary code with the whole execution configurations into a disk image, during a remote attestation. 
A problem has been found in Graphene-SGX, which does check the integrity of all loaded libraries pointed by environment variables, but fails to do so on loaded configuration files such as those under the `/etc' folder, which could expose an attack surface. We found that Chancel, Deflection, and Ratel have no such protection at all.

\vspace{2pt}\noindent$\bullet$\textit{ Mitigation}. All such meta-data should be both integrity and confidentiality protected. Also for the environment variables whose correct content cannot be determined before being loaded into the enclave, Tcons should perform security sanitization on them.  \looseness=-1


\subsection{Implications to Real-World Applications}


Inadequate protection implemented by today's Tcons, together with incomplete information about their security guarantees, could make it challenging to use them safely. To understand this challenge, we looked for clues from published research on development of Tcon-based applications~\cite{bailleu2019speicher}, ranging from IDS~\cite{kuvaiskii2018snort}, to network gateway~\cite{schwarz2020seng} and to CDNs~\cite{herwig2020achieving}. Even from the research outcomes of TEE experts, we found the signs of the difficulty in building secure applications on such less protected and ill documented Tcon designs, indicating the risks less knowledgeable users may face when using these containers. Following we describe two examples.    


\vspace{3pt}\noindent\textbf{Untrusted timers}. Prior research~\cite{kuvaiskii2018snort} reports porting Snort IDS into Graphene-SGX with only 27 Lines of Code (LoC) modified in Snort and 178 LoC in Graphene-SGX.
A problem is that the IDS heavily relies on time for detection~\cite{kuchler2021does}, invoking the syscall \verb|clock_gettime| at least twice when inspecting a single packet. This syscall is completely unmediated by Graphene-SGX, as discovered in our research. So not only is the timing information completely untrusted (which has never been made clear by the Graphene document), but the return value itself has not been properly vetted by the container for attack payloads, which exposes the IDS to wrong information and even exploits like the Iago attack.  

The authors developing the system apparently realized that the timing information is unreliable. So they built a helper “clock” thread in Graphene-SGX that roughly estimates the time. This approach, however, turns out to be fragile, as pointed by more recent work~\cite{huang2021aion} that software based timers within SGX could be easily manipulated by the administrator. Also, there is no evidence at all that the authors understand the Iago attack their IDS is exposed to.


\vspace{3pt}\noindent\textbf{Inadequate interface protection}. Another example is SENG, a shielded network gateway running in Graphene-SGX~\cite{schwarz2020seng} that enables firewalls to reliably attribute traffic to an application.
This application, however, requires a reliable domain resolving service, which the local host is not trusted to provide. To solve this problem, the authors assume a trusted DNS resolver located outside. To use the resolver, however, the in-enclave firewall needs to securely query the resolver using integrity-protected DNS variants, e.g., DNSSEC, DNS over TLS (DoT) or DNS over HTTPS. Also the enclave needs to import configuration files such as “resolv.conf”, which is assumed to be trusted. However, Graphene-SGX does not support secure communication, so the authors have to build up their own primitives for the secure query. Further our research shows that Graphene does not perform any loading time checks, so a malicious configuration file can only be captured by the host itself and could even cause an exploit before that when the loader is vulnerable. On the other hand, we found that Occlum does support loading time inspection, and can therefore check the integrity of the files and ensure their proper formats.

\section{Lesson Learned}
\label{sec:tradeoffs}


Our research reveals the inadequate protection implemented by today's Tcons and more importantly, the lack of information about what they can protect and what they cannot. Timely and complete documentation about a Tcon's capability and limitation can facilitate its real-world use, while incomplete information could expose its user to the security risk that can be avoided. Serving this purpose, we released TECUZZER~\cite{code_release}, which helps better understand the protection enforced by a Tcon, even in the absence of its source code. 

Also, when it comes to the lessons learnt for Tcon design and implementation, following are our key takeaways:

\vspace{3pt}\noindent\textbf{Ocall interface design}.
Tcon developers should narrow the syscall interfaces as much as possible, not only by reducing the number of exposed syscalls, but also through sanitizing call parameters. 
For the syscalls that need to be handled by the OS, Tcons should be able to audit them, or enable their users to configure the syscall interfaces, removing the exposed calls their applications are unlikely to use.  If none of these can be done, Tcon at lease should communicate to the user those exposed syscalls and their potential risks. \looseness=-1



\vspace{3pt}\noindent\textbf{Implementation of SFI}. SFI is critical for in-enclave isolation but it is hard to be done right, as confirmed by our conversations with Occlum's developers. Complete enforcement of SFI could incur heavy overheads. We believe that use of a static verification to reduce the need for runtime checks, as adopted by Occlum, should be the right solution. When it comes to highly pervasive instructions like direct jumps, whose inspection can even significantly impact the performance of a loading time check by the verifier, new techniques should be developed to enhance the efficiency of SFI: for example, we could perform an offline verification, using an enclave, to certify the code inspected; these programs can be directly uploaded into another enclave for an online service, after a quick checking of the certification. Another direction is to use
Webassembly, which has built-in security features, such as structured control-flow~\cite{lehmann2020everything}, and therefore is more suitable for efficient enforcement of security policies on untrusted applications. \looseness=-1

\vspace{3pt}\noindent\textbf{Side channel}. Side channels remain to be a concern that the Tcon design cannot ignore. Since a comprehensive protection can be hard, we suggest that Tcon developers should make it clear the types of side-channel threats their product can mitigate. Also a design that enables a Tcon to quickly integrate existing and newly proposed side-channel defense~\cite{shih2017t,gruss2017strong,chen2018racing,orenbach2019cosmix}, particularly those at the OS level, can be valuable for the enhancing the security quality of Tcons.

\section{Trends Today and Directions Tomorrow}








\vspace{3pt}\noindent\textbf{VM-based TEE}.
Near future TEE platforms such as ARM CCA~\cite{arm_cca_whitepaper} and Intel TDX~\cite{intel_tdx_whitepaper} can provide a virtual machine-level trusted execution environment.
Although we can foresee that these VM-based TEEs can provide the ultimate compatibility of upper layer applications, container technology is still needed on these platforms, due to its high portability, easy deployment, and lightweight.
Just as Kata container is currently trying to migrate itself to SEV, other TEE containers also can be applied to the new platforms.

As an lightweight OS, Tcons on virtual machine-based TEE should shield the underlying hardware differences and provide a consistent API on the upper layer. At present, Tcons for SGX cannot do this since they are tailored for SGX's ISAs.
Besides, Tcon developers need to consider what APIs are needed to allow an application run inside a VM domain and how to design inner isolation mechanism in different threat models. Security risks from interaction between the container and the untrusted hypervisor still exist.
How a client perform the RA, deploy, calculate, get sealed results directly using APIs without caring about the details are issues need to be taken into account. \looseness=-1


\vspace{3pt}\noindent\textbf{Tcon Management Framework}.
TEE hardwares unlock their potential when they come to server side: Intel SGX allows up to 1TB EPC memory on 3rd Scalable Xeon CPUs, and AMD SEV is available on some EPYC server CPUs. These two product lines are both designed for server. Cloud service providers also have products with TEE support ready for deployment, such as Azure Kubernetes Service~\cite{aks_website}, AWS Nitro Enclaves~\cite{aws_nitro_website}, and Alibaba Cloud~\cite{alibaba_cloud_website}.

With the inclination of TEE hardwares to the server and cloud, some TEE middlewares are migrating to cloud in recent years. Various frameworks are (or will be) able to deploy scalable confidential computing tasks in end-to-end fashion. Teaclave~\cite{teaclave_official}, Project Oak~\cite{oak_repo}, Veracruz~\cite{veracruz_repo}, Confidential Consortium Framework~\cite{russinovich2019ccf, ccf_repo}, and Marblerun~\cite{marblerun_official} features multi-party computation, blockchain, service mesh and/or all-way encryption. They are all open source projects and built on top of the aforementioned SDKs or Tcons, but these frameworks are beyond Tcons. \looseness=-1




\section{Related Work}\label{sec:relatedwork}

\vspace{3pt}\noindent\textbf{Survey on TEE}.
Jo Van Bulck et al.~\cite{van2017sgx} discuss API/ABI level sanitization vulnerabilities in Graphene and some SDK-type shielding runtimes on SGX, TrustZone, and RISC-V. They analyze the bridge functions with vulnerabilities that can lead to exploitable memory safety and side-channel issues.
Göttel et al.~\cite{gottel2018security} evaluated the impact of the memory protection
mechanisms (AMD SEV and Intel SGX) on performance and provided several trade-offs in terms of latency, throughput, processing time and energy requirements.
Cerdeira et al.~\cite{cerdeira2020sok} present a security analysis of popular
TrustZone-assisted TEE systems on commercial ARM platforms, and identified several critical vulnerabilities across existing systems.
None of the above surveys systematically investigated a new application scenario, namely Tcon, which will be a new direction of using TEE.\looseness=-1

\vspace{3pt}\noindent\textbf{Fuzzers}.
Cui et al.~\cite{cui2021emilia} developed \textit{Emilia} to automatically detect Iago vulnerabilities in legacy applications by fuzzing applications using system call return values.
Unlike Emilia, our written-in-Rust fuzzer consists of two parts, of which the kernel part is also active.
Khandaker et al.~\cite{khandaker2020coin} introduced the COIN attacks and proposed an extensible framework to test if an enclave is vulnerable such attacks with instruction emulation and concolic execution.
Cloosters et al.~\cite{cloosters2020teerex} developed TeeREX to automatically analyze enclave binary code for vulnerabilities
introduced at the host-to-enclave boundary by means of symbolic execution.
Unlike their methods to discover vulnerabilities in SGX programs, our solution is meant to fuzz the TEE containers, and does not rely on knowing source code.

\section{Conclusion}
\label{sec:conclusion}

In this paper, we provided an systematic assessment of existing TEE containers, and identified common pitfalls that are prevalent in their designs and implementations. Based on the findings, we provided recommendations to help the developers and users in avoiding them.
To improve the secure development of Tcons, a system call fuzzer and a suite of security benchmarks were proposed and implemented that are applicable to test those Tcons. We further summarized the takeaways and encourage the community to put more attention on new techniques for today's Tcons and future designs.





\end{document}